\documentclass[journal]{IEEEtran}
\usepackage{cite}

\ifCLASSINFOpdf
   \usepackage[pdftex]{graphicx}
\else
\fi
\usepackage[cmex10]{amsmath}
\usepackage{algorithmic}
\usepackage{algorithm}

\usepackage{array}

\usepackage[tight,footnotesize]{subfigure}
\newcommand{\qed}{\nobreak \ifvmode \relax \else
      \ifdim\lastskip<1.5em \hskip-\lastskip
      \hskip1.5em plus0em minus0.5em \fi \nobreak
      \vrule height0.75em width0.5em depth0.25em\fi}

\begin{document}
%
\title{Modeling Massive Amount of Experimental Data with Large Random Matrices in a Real-Time UWB-MIMO System}
%
%
%

\author{Xia Li,~\IEEEmembership{Student Member,~IEEE,}
        Feng Lin,~\IEEEmembership{Student Member,~IEEE,}
        Robert C. Qiu,~\IEEEmembership{Senior Member,~IEEE}
\thanks{X. Li is currently with Qualcomm Incorporated, San Diego, CA. X. Li was with the Department of Electrical and Computer Engineering, Tennessee Technological University, Cookeville, TN, 38505, e-mail:sharpli.uestc@gmail.com}
\thanks{F. Lin and R. C. Qiu are with the Department of Electrical and Computer Engineering, Center for Manufacturing Research, Tennessee Technological University, Cookeville, TN, 38505, e-mail: fenglin@ieee.org; rqiu@ieee.org.}
}

\maketitle

\begin{abstract}
The aim of this paper is to study data modeling for massive datasets. Large random matrices are used to model the massive amount of data collected from our experimental testbed. This testbed was developed for a real-time ultra-wideband, multiple input multiple output (UWB-MIMO) system. Empirical spectral density is the relevant information we seek for. After we treat this UWB-MIMO system as a black box, we aim to model the output of the black box as a large statistical system, whose outputs can be described by (large) random matrices. This model is extremely general to allow for the study of non-linear and non-Gaussian phenomenon. The good agreements between the theoretical predictions and the empirical findings validate the correctness of the our suggested data model.

\end{abstract}


%
\IEEEpeerreviewmaketitle

\section{Introduction}
\label{sec_intro}
The purpose of statistical methods is to reduce a large quantity of data to a few that are capable of containing as much as possible of the relevant information in the original data. Because the data will generally supply a large number of ``facts'', many more than are sought, much information in the data is irrelevant.  In this paper, our relevant information is the empirical spectral density of the (random) data matrix.

In~\cite{QiuBook2012Cognitive}, the idea of modeling Big Data with (large) random matrices was proposed for the first time, where the theory for large random matrices may be, arguably, the foundation for communications, networking and sensing. In  recent works~\cite{Qiu_WicksBook2013,QiuAntonik2014Wiley}, this idea has been carried out from distributed sensing to the Smart Grid. It is argued in~\cite{QiuAntonik2014Wiley} that the theme of the 5G wireless network may be unified in the framework of Big Data.

These series works~\cite{QiuBook2012Cognitive,Qiu_WicksBook2013,QiuAntonik2014Wiley}   clearly articulated the ideas and treated the necessary analytical tools. The concrete results, on the other hand, are still few,---maybe none, to our best knowledge---especially using experimental data. This short paper is aimed at filling this gap. In particular, we deal with the pressing issue daily encountered in our Lab: to make sense of the massive amount data that are accumulated at an accelerating speed. The UWB-MIMO system~\cite{li2012,li2012waveform} is used as an experimental testbed. After we treat this UWB-MIMO system as a black box,---the assumption of linear time-varying system can be discarded---our goal is to model the output of the black box as a large statistical system, whose outputs can be described by  (large) random matrices. This model is extremely general to allow for the study of non-linear and non-Gaussian phenomenon. 

The structure of this paper is as follows. First, the theoretical models are introduced. Then, the empirical spectral results are compared with the theoretical predictions. The good agreements validate the correctness of the our suggested data model.
\section{Empirical Spectral Density of Random Matrix}
\label{sec_ker}

Suppose that $X_{ij}$ are independent and identically distributed (i.i.d.) real random variables. Let $\mathbf{X}_{n} = \left (X_{ij}  \right )_{p\times n}$ and $\mathbf{T}_{n}$ be a $p\times p$ nonrandom Hermitian nonnegative definite matrix. Consider the random matrices~\cite{jing2010nonparametric}
\begin{equation}
\label{A_n}
\mathbf{A}_{n} = \frac{1}{n}\mathbf{T}_{n}^{1/2}\mathbf{X}_{n}\mathbf{X}_{n}^{T} \mathbf{T}_{n}^{1/2}.
\end{equation}
When $E\left ( X_{i,j} \right ) = 0$ and $E\left ( X_{i,j}^{2} \right ) = 1$, $\mathbf{A}_{n}$ can be viewed as a sample covariance matrix drawn from the population with covariance matrix $\mathbf{T}_{n}$. Moreover, if $\mathbf{T}_{n}$ is another sample covariance matrix, independent of $\mathbf{X}_{n}$, then $\mathbf{A}_{n}$ is a Wishart matrix.

The basic limit theorem regarding $\mathbf{A}_{n}$ concerns its empirical spectral distribution $F^{\mathbf{A}_{n}}$, which is given by
\begin{equation}
\label{F_A}
F^{\mathbf{A}_{n}}\left ( x \right )=\frac{1}{p}\sum_{p}^{k=1}I\left ( \lambda _{k} \leq x\right ),
\end{equation}
where $\lambda _{k}, k = 1, \cdots, p$ denote the eigenvalues of $\mathbf{A}_{n}$.

Suppose the ratio of the dimension to the sample size $c_{n} = p/n$ tends to $c$ as $n\rightarrow \infty $. When $\mathbf{T}_{n}$ becomes the identity matrix, $F^{\mathbf{A}_{n}}$ tends to the well-known Marcenko and Pastur law with the density function
\begin{equation}
\label{MP}
f_{c}\left ( x \right )= \left\{\begin{matrix}
\left ( 2\pi cx \right )^{-1}\sqrt{\left ( b-x \right )\left ( x-a \right )},\; a\leq x\leq b,\\ 
0,\;\;\;\;\;\;\;\;\;\;\;\;\;\;\;\;\;\;\;\;\;\;\;\;\;\;\;\;\;\;\;\;\;\;\;\;\;\;\; \mathrm{otherwise}.
\end{matrix}\right.
\end{equation}
It has point mass $1-c^{-1}$ at the origin if $c > 1$, where $a = \left ( 1 - \sqrt{c} \right )^{2}$ and $b = \left ( 1 + \sqrt{c} \right )^{2}$.

Suppose $f\left ( x \right )$ is an unknown density function of $X_{i,j}$, and $F_{n}\left ( x \right )$ is the empirical distribution function determined by the sample. A popular nonparametric estimate of $f\left ( x \right )$ is then
\begin{equation}
\label{kernel_equ}
\begin{array}{*{20}{l}}
{\hat{f_n}\left( x \right) = \frac{1}{{nh}}\sum\limits_{j = 1}^n K \left( {\frac{{x - {X_j}}}{h}} \right)}\\
{\quad {\mkern 1mu} {\mkern 1mu} {\mkern 1mu} {\mkern 1mu} \quad {\mkern 1mu}  = \frac{1}{h}\int K \left( {\frac{{x - y}}{h}} \right)d{F_n}\left( y \right)}
\end{array}
\end{equation}
where the function $K(\cdot)$ is a Borel function and $h = h\left ( n \right )$ is the bandwidth which tends to 0 as $n\rightarrow \infty$. 


Under some regularity conditions on the kernel function $K(\cdot),$ it is well known that $\hat{f_{n}}\left ( x \right )\rightarrow f\left ( x \right )$ in some sense. The kernel density estimation would be
\begin{equation}
\label{kernel_equ2}
\begin{array}{l}
{f_n}\left( x \right) = \frac{1}{{ph}}\sum\limits_{i = 1}^p K \left( {\frac{{x - {\mu _i}}}{h}} \right)\\
\quad \, \, \, \, \quad\, = \frac{1}{h}\int K \left( {\frac{{x - y}}{h}} \right)d{F^{{{\bf{A}}_n}}}\left( y \right)
\end{array}
\end{equation}
where $\mu _{i}, i =1,\cdots, p$, are eigenvalues of $\mathbf{A}_{n}$.
\section{Eigenvalue Spectra of Time-lagged Correlation Matrices}
\label{timelag}

The time-lagged analog to the sample covariance matrix is defined as $C_{\tau }^{ij}\sim \sum_{t}^{T}r_{t}^{i}r_{t-\tau }^{j}$, where one time series is shifted by $\tau$ timesteps with respect to the other. Sample covariance matrix can be viewed as a special case of time-lagged covariance matrix with $\tau = 0$, also known as equal-time correlations. For $\tau \neq 0$, the lagged correlation matrix $\mathbf{C}_{\tau } $ is \textit{non-symmetric}\footnote{Thus its eigenvalues are complex numbers, in contrast with the real values of symmetric sample covariance matrix.} and contains the lagged auto-correlations in the diagonal. It can be written as
\begin{equation}
\label{C_T}
\mathbf{C}_{\tau } = \frac{1}{T}\mathbf{X}\mathbf{D}_{\tau }\mathbf{X}^{\mathrm{T}}
\end{equation}
where $\mathbf{D}_{\tau }\equiv \delta _{t,t+\tau }$ and where $\mathbf{X}$ is the $N \times T$ normalized time-series data.

The projection of eigenvalue density $\rho\left ( r \right )$ onto the $\mathit{x}$-axis is denoted by $\rho_{x}\left ( r \right )$, and the projection onto the $\mathit{y}$-axis is denoted by $\rho_{y}\left ( r \right )$. They are nothing but the rescaled spectra of the solution to the symmetric, $\rho^{S}\left ( \lambda  \right )$, and to the anti-symmetric problem, $\rho^{A}\left ( \lambda  \right )$. To be more explicit, we have~\cite{biely2008random}
\begin{equation}
\label{rou_lambda}
\begin{array}{l}
{\rho _x}\left( \lambda  \right) \equiv \rho \left( {{\mathop{\rm Re}\nolimits} \left( \lambda  \right)} \right) = \int\limits_{\rm{S}} {\rho \left( r \right)} dy = {\rho ^S}\left( {\sqrt 2 x} \right)\\
{\rho _y}\left( \lambda  \right) \equiv \rho \left( {{\mathop{\rm Im}\nolimits} \left( \lambda  \right)} \right) = \int\limits_{\rm{S}} {\rho \left( r \right)} dx = {\rho ^A}\left( {\sqrt 2 y} \right),
\end{array}
\end{equation}
where the integration extends over the support $\rm{S}$ in the complex plane. 

The eigenvalue density of the symmetric problem can be obtained from the well-known relation
\begin{equation}
\label{rou_s}
\rho ^{S}\left ( x \right )= \sum_{n}\delta \left ( x-x_{n} \right )= \frac{1}{\pi }\lim_{\epsilon \rightarrow 0}\left [ \textrm{Im}\left ( G^{S}\left ( x-i\epsilon  \right ) \right ) \right ].
\end{equation}
For a radial symmetric problem, $\rho ^{S}\sim \rho ^{A}$. The $G(z)$ in~\eqref{rou_s} is the Greens function given by
\begin{equation}
\label{Green_f}
\begin{array}{l}
\frac{1}{{{Q^3}}}{z^2}{G^4}\left( z \right) - 2\frac{1}{{{Q^2}}}\left( {\frac{1}{Q} - 1} \right)z{G^3}\left( z \right)\\
 - \frac{1}{Q}\left( {{z^2} - {{\left( {\frac{1}{Q} - 1} \right)}^2}} \right){G^2}\left( z \right)\\
 + 2\left( {\frac{1}{Q} - 1} \right)zG\left( z \right) + 2 - \frac{1}{Q} = 0,
\end{array}
\end{equation} 
with $Q\equiv T/N$ playing the role of a information-to-noise ratio. This equation is independent of a specific value for $\tau$ and is valid for any value of it.

\section{Experimental Results of Empirical Spectral Density}
\label{sec_result}


\subsection{Data Collection Using a Real-Time UWB MIMO System}

\begin{figure}[!t]
	\centering
	\includegraphics[width=3.0in]{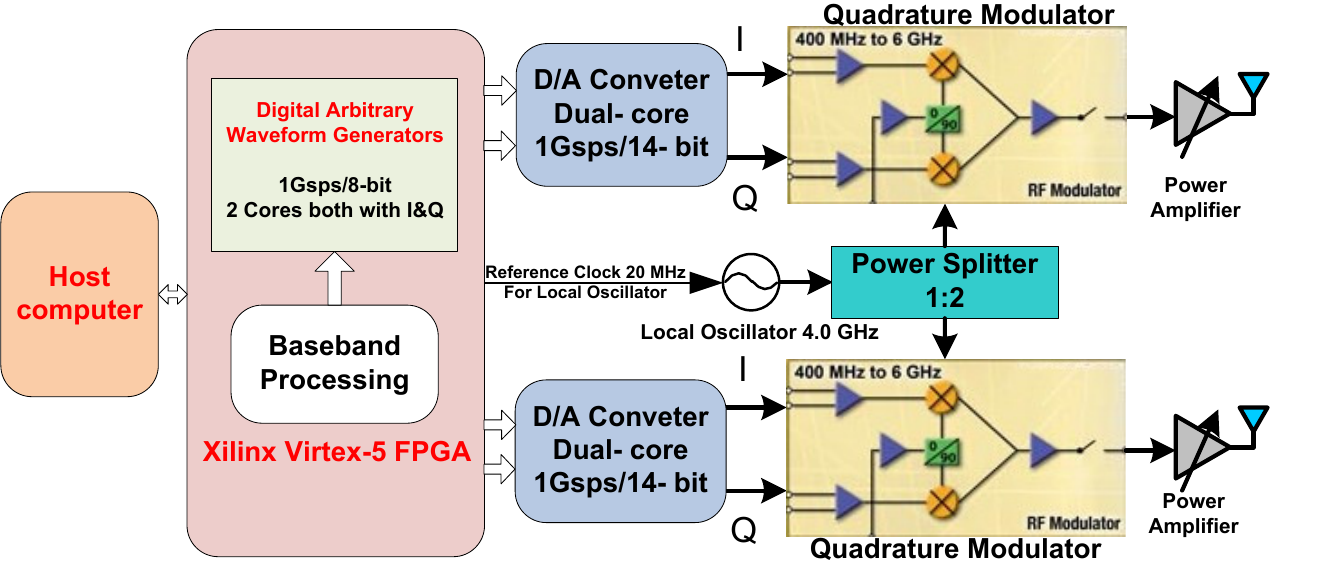}
	\caption{Schematic of transmitter design.}
	\label{transmitter}
\end{figure}

The transmitter architecture is the same as the previous work~\cite{song2012real}, as shown in Fig.~\ref{transmitter}. The carrier frequency of the system is controlled by the oscillator in the transmitter. In the experiment, the carrier frequency is set to 4 GHz. The bandwidth of the radar waveform is 500 MHz, making the system an ultra-wideband system. 

\subsection{Receiver Design for Compressive Sensing Based Radar}

\begin{figure}[!t]
	\centering
	\includegraphics[width=3.0in]{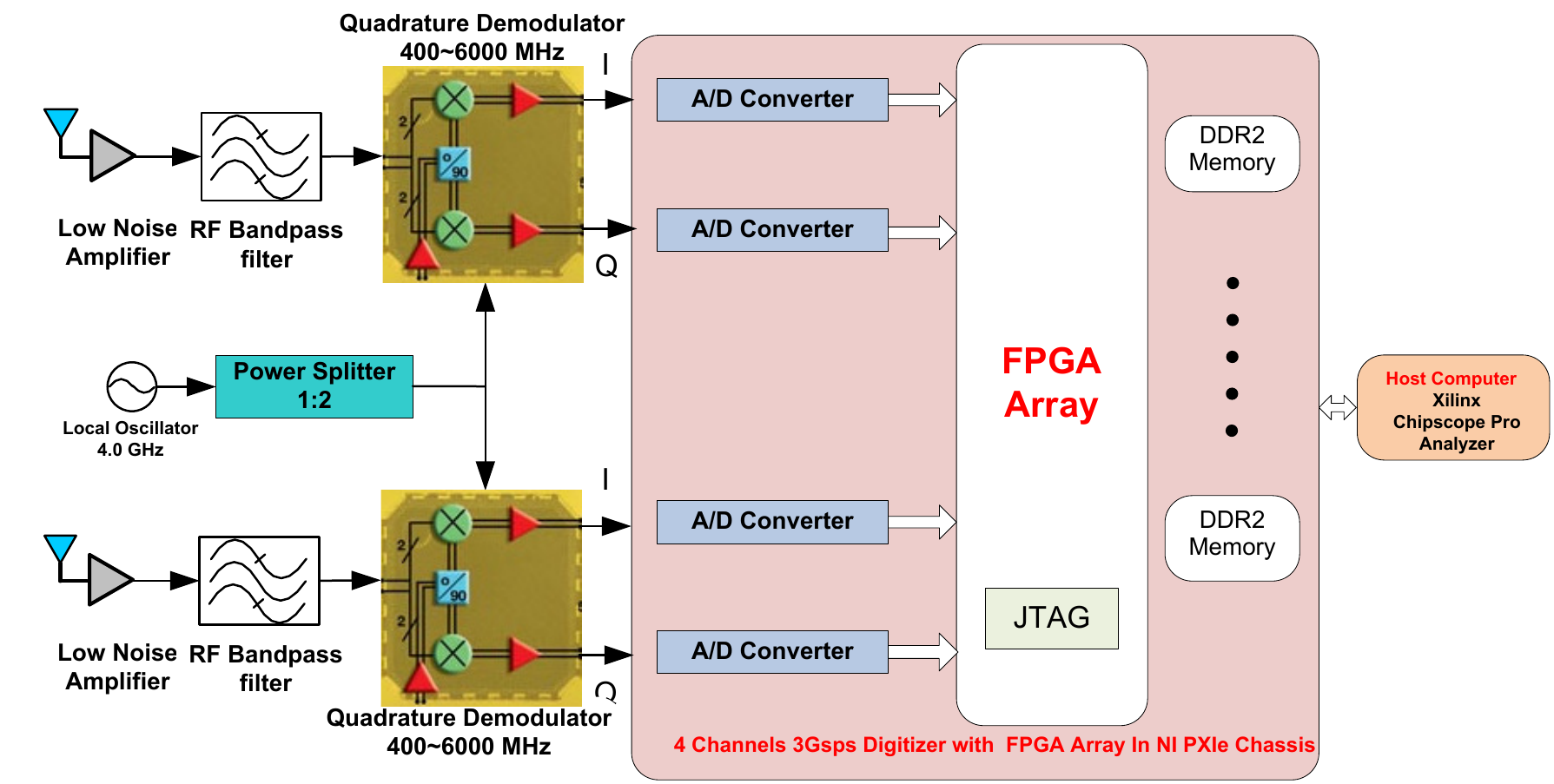}
	\caption{Schematic of compressive sensing receiver design.}
	\label{receiver}
\end{figure}

To recover the channel information from the received radar waveform, in traditional system, the sampling rate of the received radar waveform must be higher than 500 MHz. For the testbed shown in Fig.~\ref{receiver}, the ADC is working at 3 Gsps. 
The sampling rate of the ADC is changeable and controlled by a local oscillator. The analog RF down-converter is the same as the traditional system. A bandpass filter is used to remove the noise outside of radar waveform bandwidth. 




\subsection{Experimental Results}
For a data matrix $\bf X$ of $n\times p,$   we have  $n = 1024$ and  $p = 4096,$ with $c _{n} = p/n=4 $ accordingly. 

\begin{figure}[!t]
\centering
\includegraphics[width=3.0in]{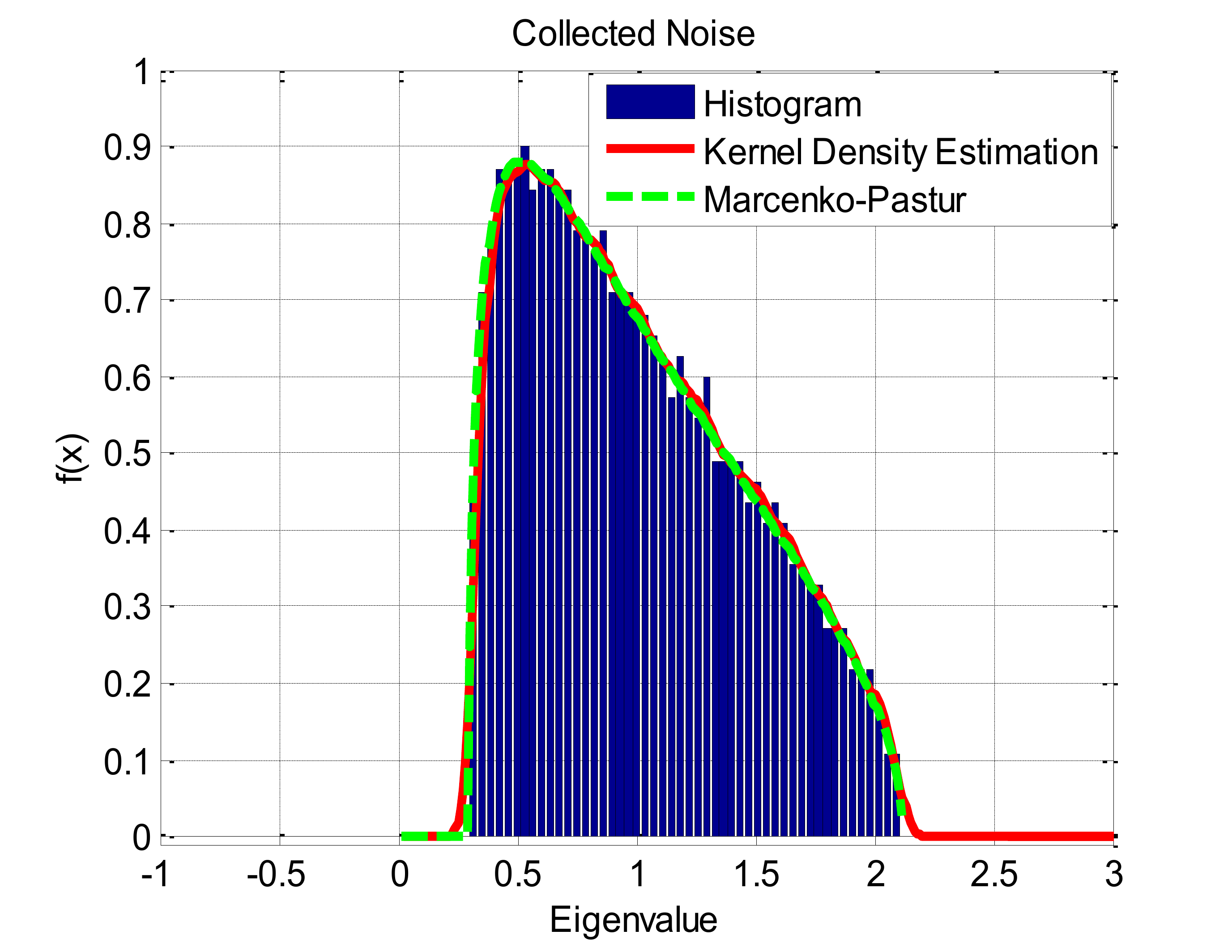}
\caption{Comparison of empirical spectral density with Marcenko-Pastur law using data  consisting of white Gaussian noise only.}
\label{noise_ker}
\end{figure}

\textit{Case 1}: No signal is transmitted at the transmitter, only  noise is present at the receiver. Because white Gaussian noise is of i.i.d. nature, its eigenvalues distribution exactly follows Marcenko-Pastur law~\eqref{MP}. This empirical agreement with the theoretical predictions is shown in Fig.~\ref{noise_ker}. 

\begin{figure}[!t]
\centering
\includegraphics[width=3.0in]{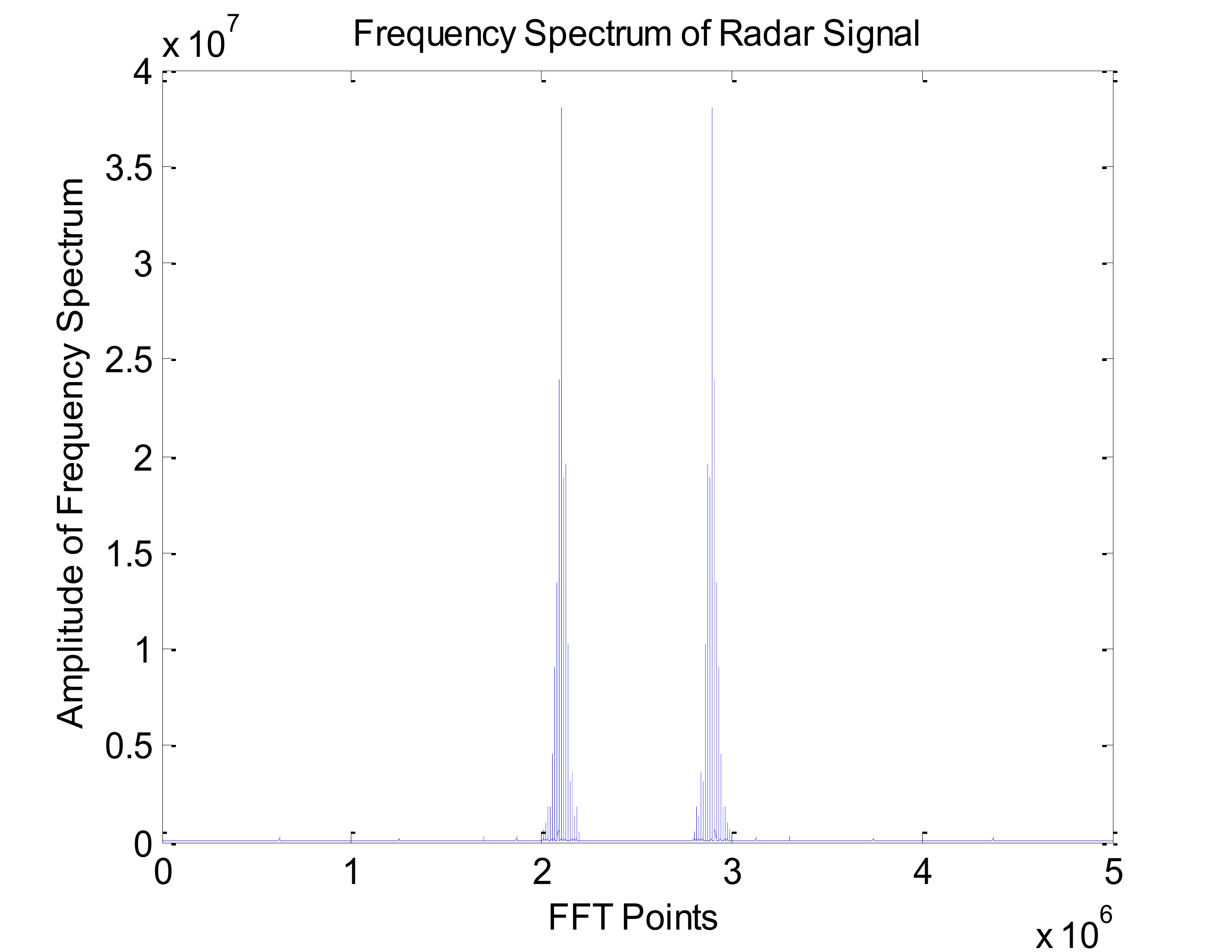}
\caption{Frequency spectrum of narrowband modulation signals.}
\label{radar_spectrum}
\end{figure}

\begin{figure}[!t]
\centering
\includegraphics[width=3.0in]{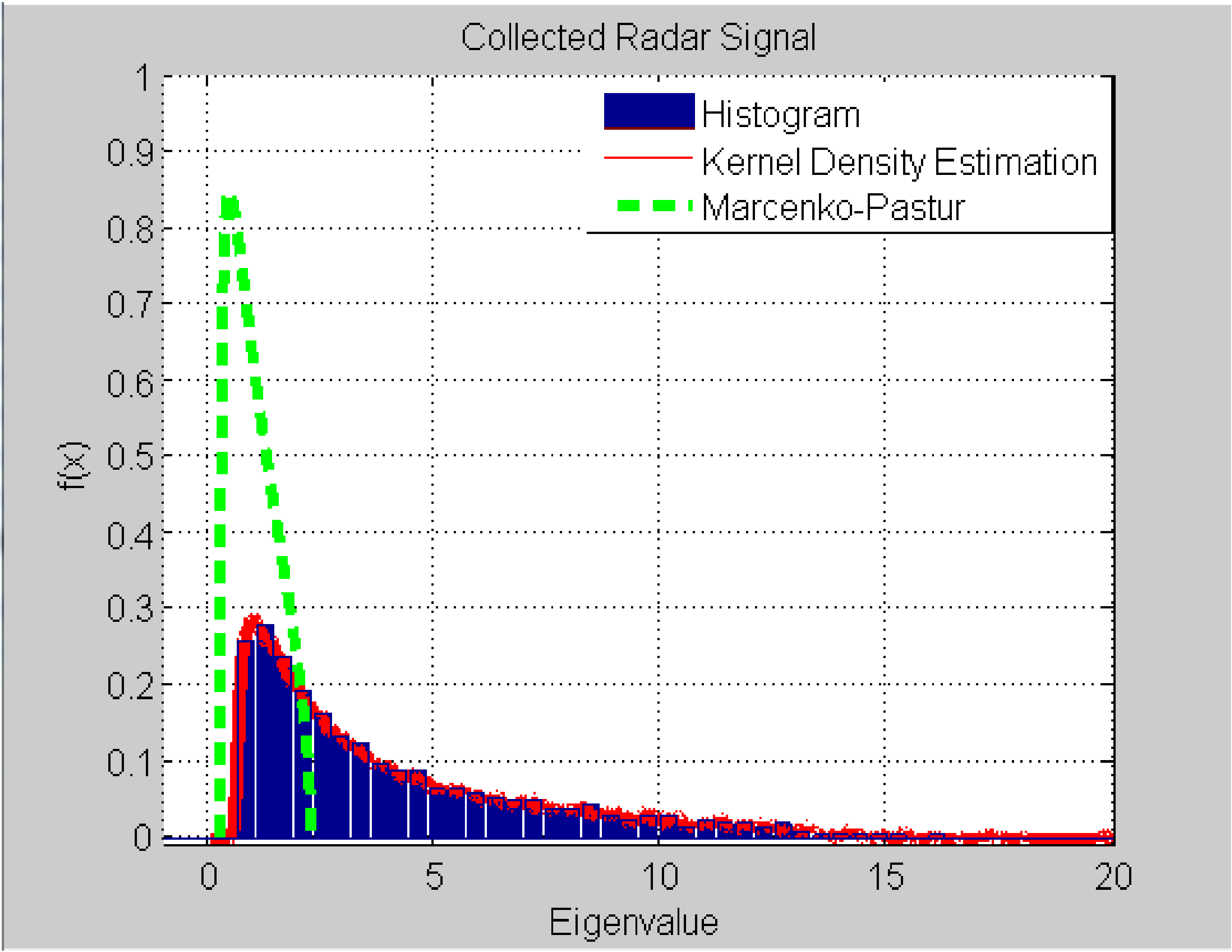}
\caption{Comparison of empirical spectral density with Marcenko-Pastur law using data  consisting of narrowband modulation waveforms (see Fig.~\ref{radar_spectrum}). }
\label{radar_ker}
\end{figure}

\textit{Case 2}: Narrow band radio signal (Fig.~\ref{radar_spectrum}) is transmitted and received through the air at a distance. The result is shown in Fig.~\ref{radar_ker}.\footnote{Because the data for the histogram drawing is too big for MATLAB to execute the copy figure command, Fig.~\ref{radar_ker} can  be only obtained by printing screen.} Because the received radar signal is highly correlated, the empirical spectral density deviates from the Marcenko-Pastur law, as expected from the random matrix theory. 

\begin{figure}[!t]
\centering
\includegraphics[width=3.0in]{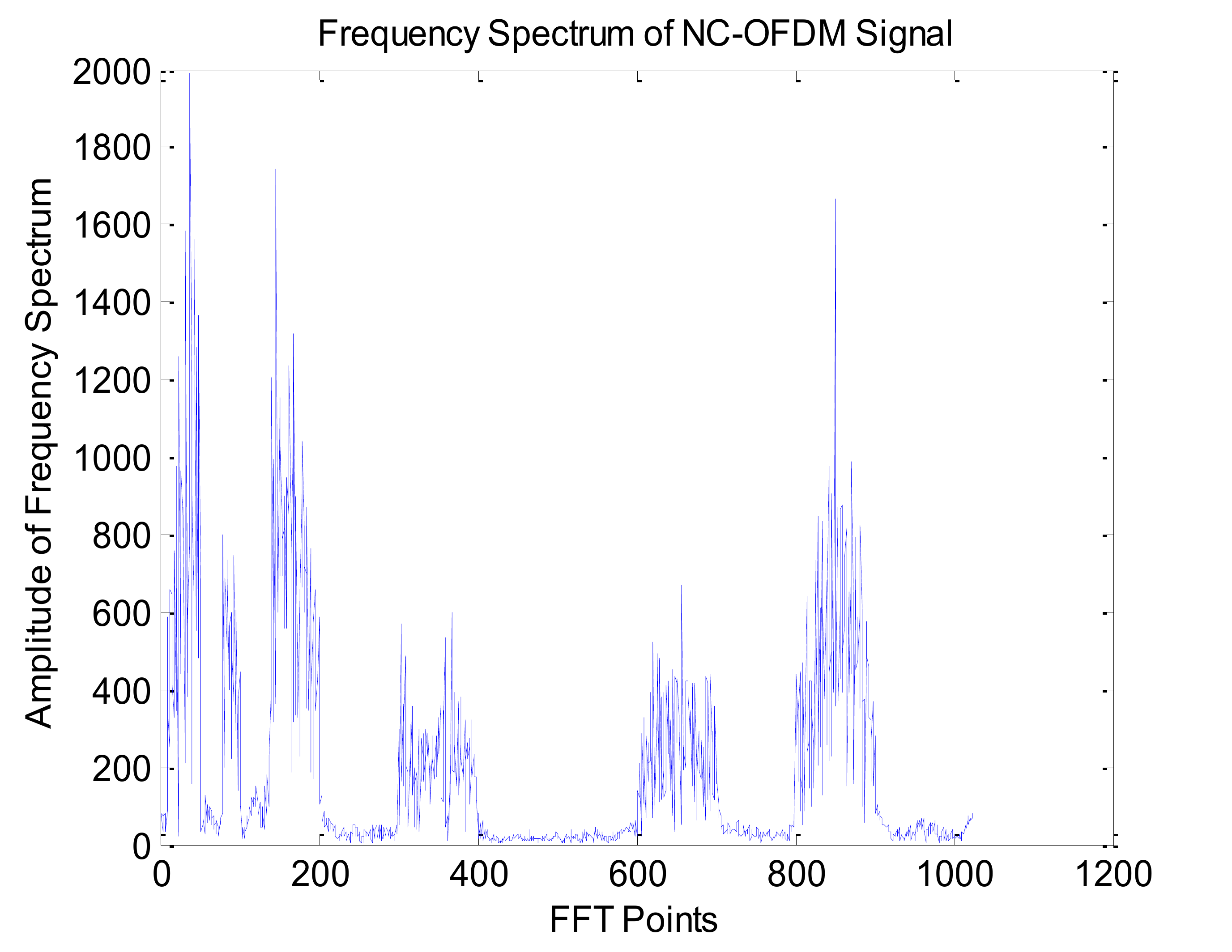}
\caption{Frequency spectrum of NC-OFDM signal with 426 out of 1024 sub-carriers.}
\label{ofdm_spectrum}
\end{figure}

\begin{figure}[!t]
\centering
\includegraphics[width=3.0in]{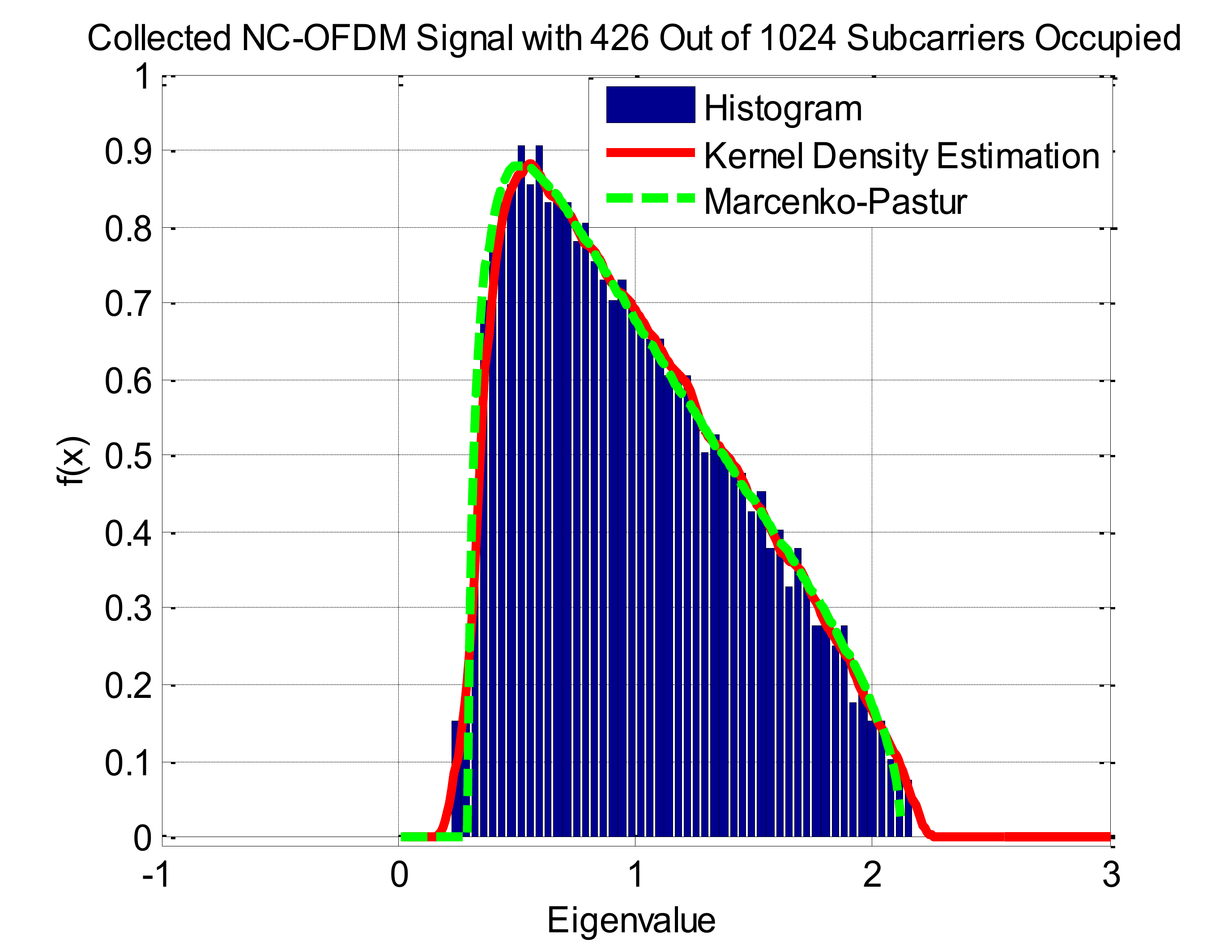}
\caption{Comparison of empirical spectral density with Marcenko-Pastur law using data  consisting of NC-OFDM modulation waveforms.  Only 426 tones (out of 1024 tones) are used, as shown in Fig.~\ref{ofdm_ker}.}
\label{ofdm_ker}
\end{figure}

\textit{Case 3}: Wide band NC-OFDM signal is used as transmission signal with BPSK baseband modulation. NC-OFDM offers more flexibility compared with OFDM signal due to the capability of configuring its occupancy of sub-carriers (also called tones). The toal number of sub-carriers  of NC-OFDM signal in our experiment is 1024, in which only 426 sub-carriers are occupied. The positions of the valid sub-carriers are chosen arbitrarily. In this case, they are allocated in the positions with the indexes of $[10 : 50, 80 : 100, 140: 200, 300:400, 600:700, 800:900].$ The received time domain signal is first converted to the frequency domain by fast Fourier transform (FFT). Then the frequency domain signal is used to form a (large) random matrix $\bf X,$ to obtain its eigenvalues. The frequency spectrum of the received NC-OFDM signal is shown in Fig.~\ref{ofdm_spectrum}. The perfect agreement between the empirical spectral density and its theoretical prediction (Marcenko-Pastur law)  is illustrated in Fig.~\ref{ofdm_ker}. This is, remarkably, similar to Fig.~\ref{noise_ker}---the white Gaussian noise only. The reason behind the remarkable result is that the carriers of the frequency domain NC-OFDM signal can be modeled as i.i.d. random variables. It is natural to expect that  a result similar to Fig.~\ref{ofdm_ker} will be obtained, if all the 1024 tones are used, instead of 426 tones. This is indeed the case, validated using our experimental data.

\section{Experimental Results of Time-lagged Correlation Matrices}
\label{expe_timelag}

Any deviation from the benchmark Marcenko-Pastur law---valid for a random sample covariance matrix obtained using i.i.d. random variables---indicates the presence of some useful correlation--called ``signal''. In this section, we study the benchmark for the time-lagged correlation matrix, obtained by solving~\eqref{Green_f}.



\begin{figure}[!t]
\centering
\includegraphics[width=3.0in]{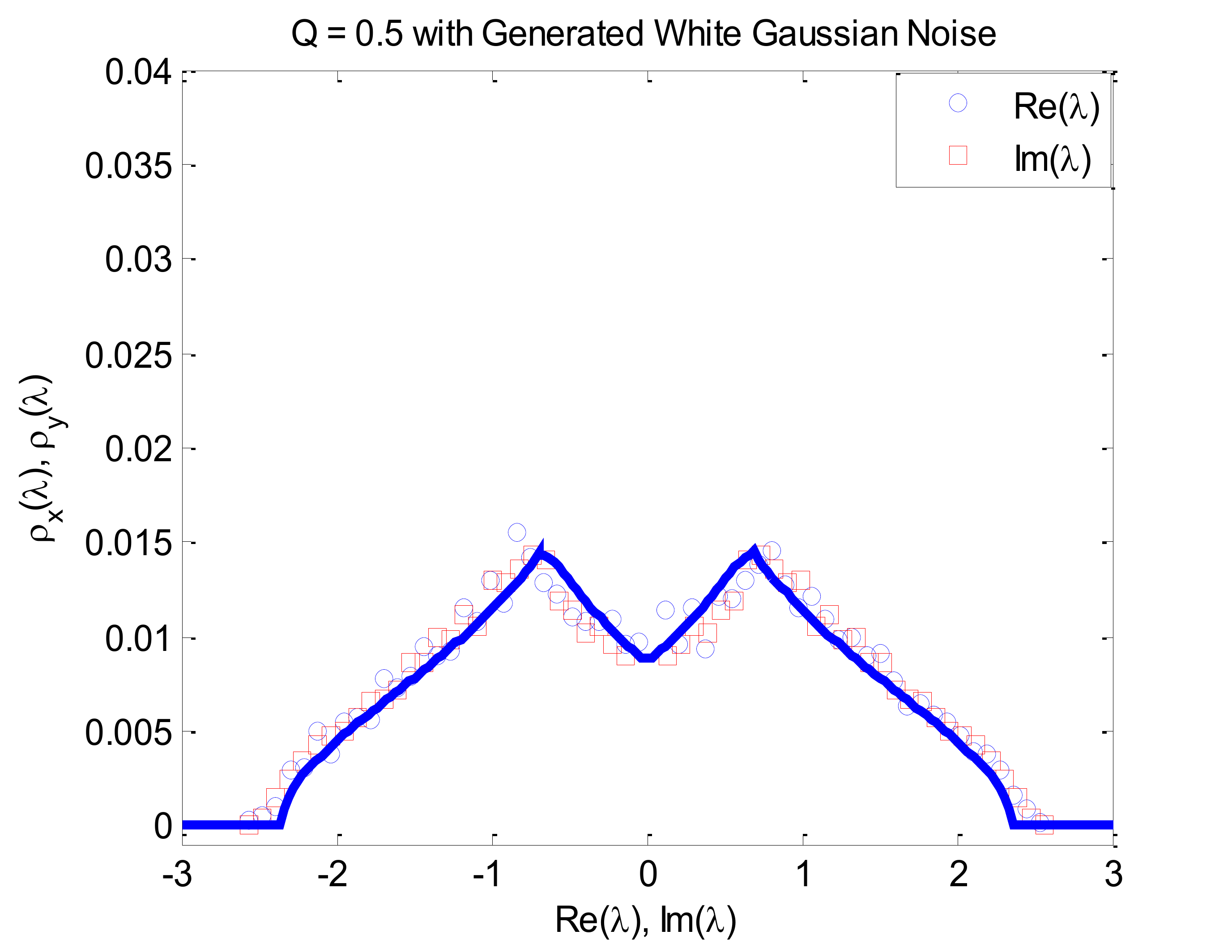}
\caption{Eigenvalue density using simulated white Gaussian noise. Q = 0.5.}
\label{Q05}
\end{figure}
\begin{figure}[!t]
\centering
\includegraphics[width=3.0in]{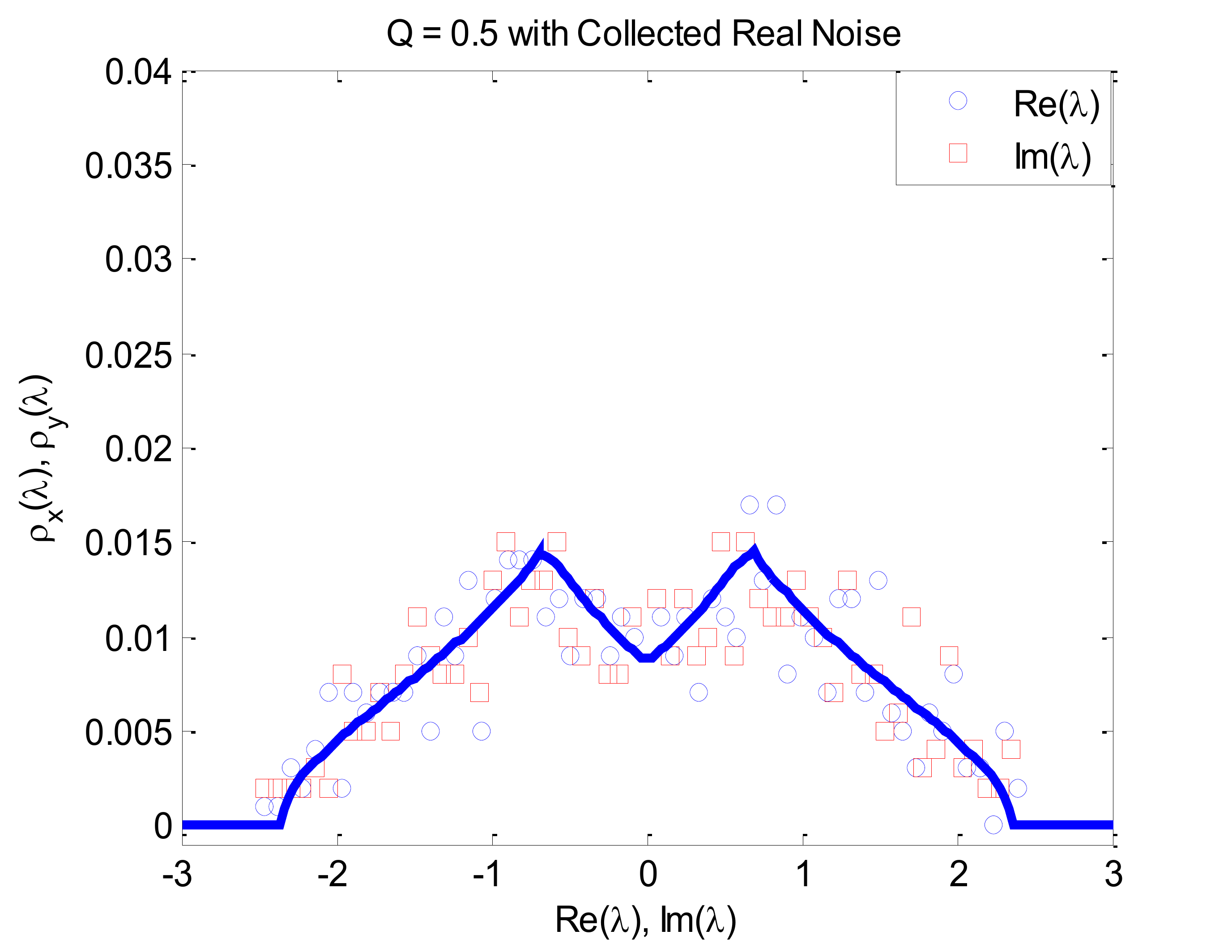}
\caption{Eigenvalue density using collected noise. Q = 0.5.}
\label{Q05_OwnNoise}
\end{figure}
\begin{figure}[!t]
\centering
\includegraphics[width=3.0in]{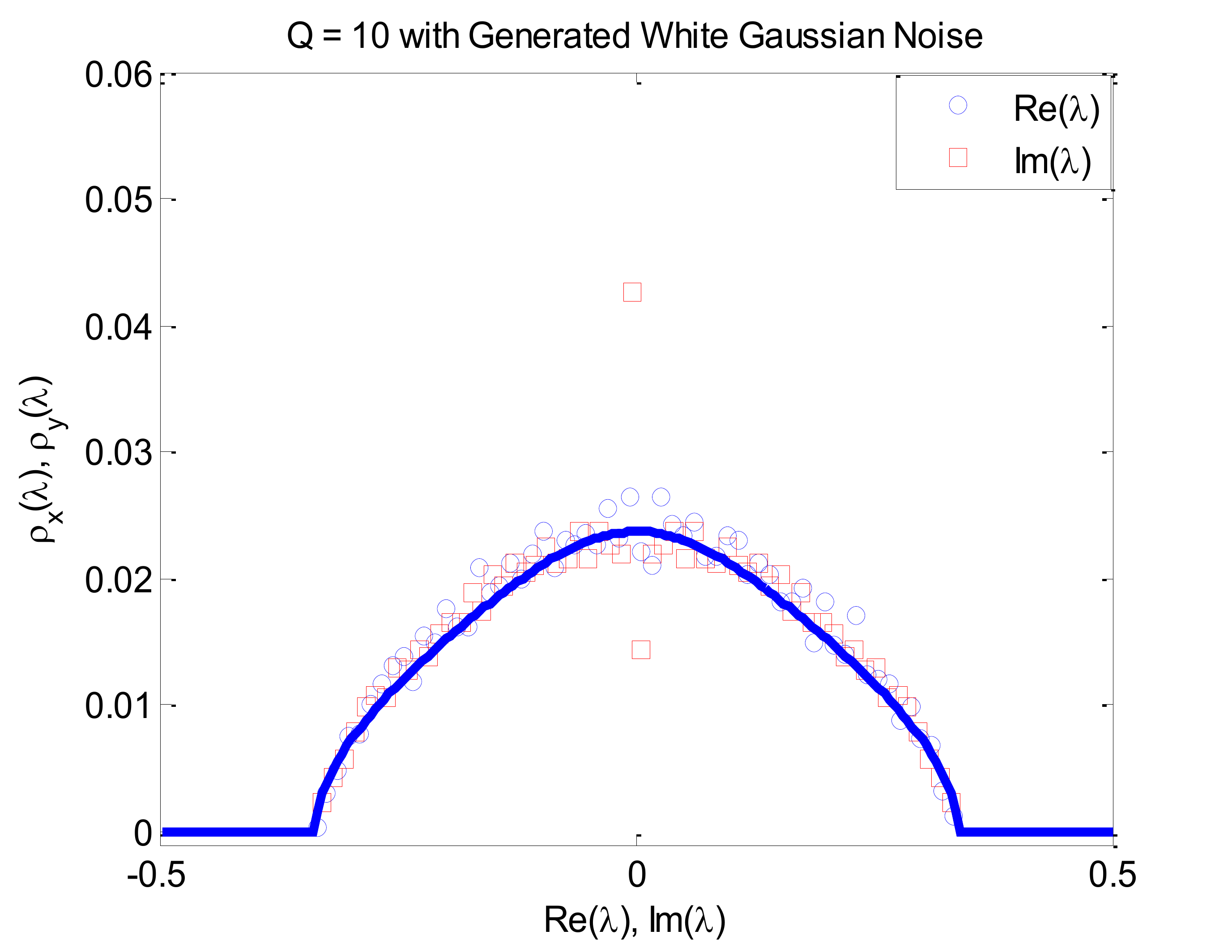}
\caption{Eigenvalue density using simulated white Gaussian noise. Q = 10.}
\label{Q10}
\end{figure}
\begin{figure}[!t]
\centering
\includegraphics[width=3.0in]{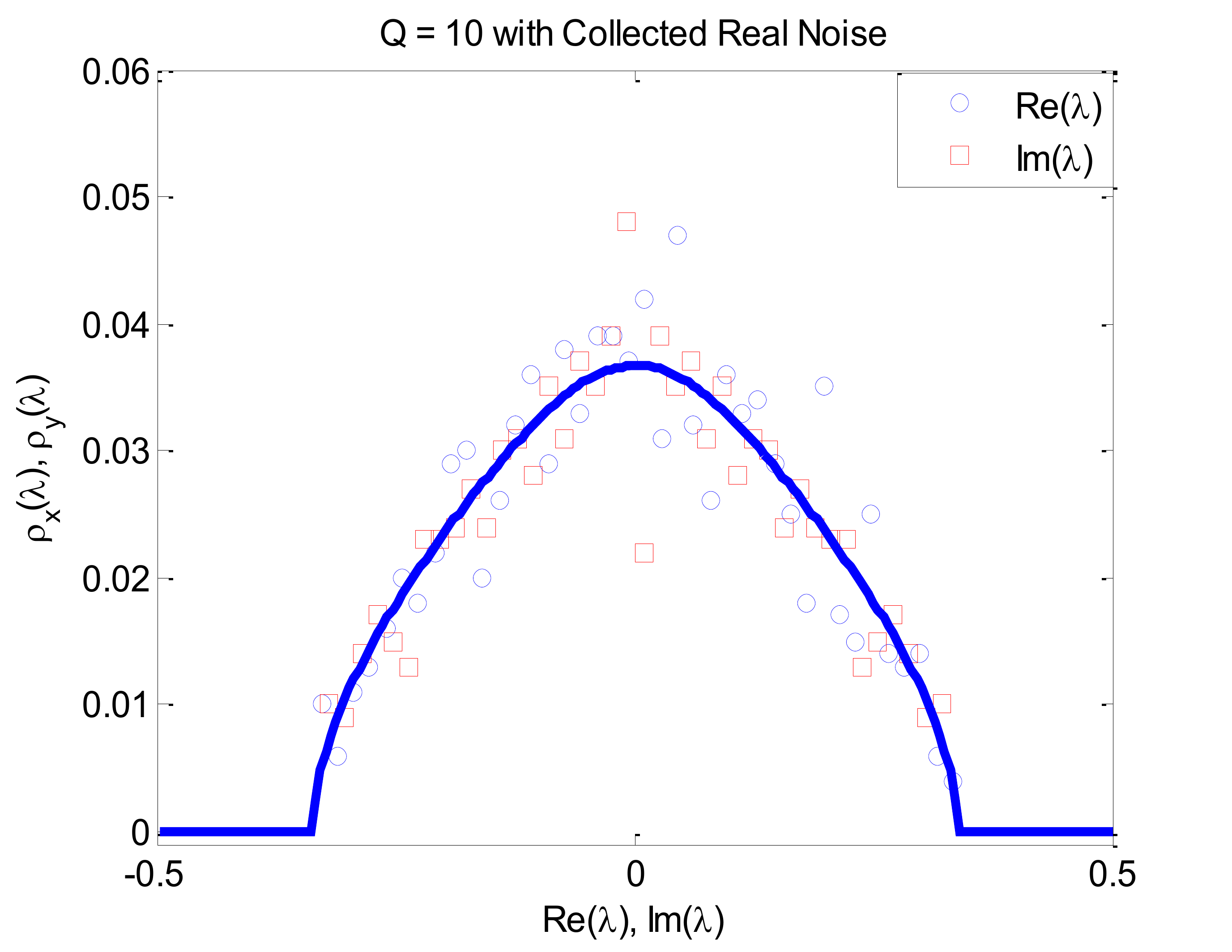}
\caption{Eigenvalue density using collected noise. Q = 10.}
\label{Q10_OwnNoise}
\end{figure}

The comparisons with $Q = 0.5$ and $Q = 10$ are shown in Fig.~\ref{Q05}, Fig.~\ref{Q05_OwnNoise} and Fig.~\ref{Q10}, Fig.~\ref{Q10_OwnNoise}, respectively. 

\section{Conclusion}
\label{conc}
Treating a ultra-wideband, multiple input mutiple output (UWB-MIMO) system testbed as a black box, we have modeled the output of the system testbed as a large statistical system, whose outputs can be described by  (large) random matrices. This model is extremely general to allow for the study of non-linear and non-Gaussian phenomenon. The good agreements between the theoretical models and the empirical results validate the correctness of the our suggested data model. This is the initial experimental demonstration of the framework of modeling massive datasets using large random matrices, which, to our best knowledge, is the first time in the field of  wireless communications and radar~\cite{QiuBook2012Cognitive,Qiu_WicksBook2013,QiuAntonik2014Wiley}.

\ifCLASSOPTIONcaptionsoff
  \newpage
\fi




    \bibliographystyle{ieeetr}   

 \begin {small}
 
 \bibliography{Bible/5GWirelessSystem,Bible/Big_Data,Bible/Compressed_Sensing,Bible/Smart_Grid,Bible/Graph_Complex_Network,Bible/Machine_Learning,Bible/Convex_Optimization,Bible/LowRankMatrixRecovery,Bible/Concentration_of_Measure,Bible/ClassicalMatrixInequalities,Bible/CompeltelyPositiveMaps,Bible/QuantumChannel,Bible/QuantumHypothesisTesting,Bible/TraceInequalities,Bible/QuantumInformation,Bible/Matrix_Inequality,Bible/RandomMatrixTheory,Bible/Fractional_Integration_bib,Bible/UWB_bib,Bible/Qiu_Group_bib,Bible/LTI_Comm_Theory_bib,Bible/Software_Defined_Radio_bib,Bible/Time_Reversal_bib,Bible/MIMO_bib,Bible/Radar_Waveform_Optim_bib,Bible/Information_Theory_bib,Bible/Compressed_Sensing_Theory,Bible/Compressed_Sensing_UWB,Bible/MISC_bib,Bible/CS_Applications,Bible/CS_Data_Stream_Algorithms,Bible/CS_Extensions,Bible/CS_Foundations_Connections,Bible/CS_Multisensor_Distributed,Bible/CS_Recovery_Algorithms,Bible/CognitiveRadio/Cognitive_Radio_bib,Bible/CognitiveRadio/CognitiveRadio2008_bib,Bible/CognitiveRadio/DySpan2007,Bible/CognitiveRadio/DySpan2005,Bible/CognitiveRadio/Gardner,Bible/CognitiveRadio/jsac200703,Bible/CognitiveRadio/jsac200801,Bible/CognitiveRadio/sensingDTV,Bible/CognitiveRadio/ucberkeley,Bible/CognitiveRadio/UWBCognitiveRadio_bib,Bible/CognitiveRadio/IEEE_JSSP_2008,Bible/CognitiveRadio/Bayesian_Network_Cognitive_Radio,Bible/CognitiveRadio/Exploiting_Historical_Spectrum,Bible/CognitiveRadio/Duke_Carin,Bible/CognitiveRadio/LiHu_upper,Bible/CognitiveRadio/AFRLref,Bible/CognitiveRadio/SVM,Bible/CognitiveRadio/NSF_ECCS_0821658,Bible/CognitiveRadio/SmartGrid,bib_xia/big_data}

\end {small}
\end{document}